\documentclass[preprint,12pt]{elsarticle}

\usepackage{amssymb}

\usepackage{booktabs} 
\usepackage{color}
\usepackage{amsfonts,amssymb,amsmath}
\usepackage{amsmath,graphicx}
\usepackage{graphicx,pifont} 
\usepackage{epsfig}
\usepackage{parallel}
\usepackage{esdiff}

\journal{Annals of Physics}

\begin{document}

\begin{frontmatter}



\title{Principles of relativistic quantum statistical thermodynamics: a class of exactly solvable models}


\author{A. Yu. Zakharov} 
\ead{anatoly.zakharov@novsu.ru}
\affiliation{organization={Dept. General and Experimental Physics,\\ Yaroslav-the-Wise Novgorod State University}, 
            addressline={41, B. Sanct-Petersburgskaya}, \\
            city={Veliky Novgorod},
            postcode={173003}, 
            country={Russian Federation}}

\begin{abstract}
A system of interacting atoms is represented as a union of two subsystems, one of which is the system of atoms, and the other is an auxiliary scalar covariant field, which is equivalent to a given static interatomic potential of general form only in the non-relativistic approximation.
It is shown that the auxiliary field is a superposition of Klein-Gordon fields, the parameters of which are related to singular points of the Fourier transform of the corresponding interatomic potential.
The general form of the relativistic Hamiltonian of a system of interacting atoms is established.
It is shown that the exact calculation of the relativistic partition function of a system of interacting atoms, taking into account the field degrees of freedom, reduces to renormalizing the parameters of the auxiliary field.
It is established that the field degrees of freedom lead to a divergence in the total energy of a classical relativistic system—an analogue of the ultraviolet catastrophe. Quantization of the auxiliary field eliminates this divergence.
The existence of a phase transition within the framework of relativistic quantum statistical thermodynamics has been proven.
	
\end{abstract}


\end{frontmatter}


\section{Introduction}

Classical Gibbs statistical mechanics aims to provide a microscopic foundation of thermodynamics~\cite{Gibbs1902},
basing on two fundamental \textit{postulates}:

\begin{enumerate}
	\item The microscopic dynamics of a system of particles (atoms) is subject to classical Newtonian mechanics, presented in Hamiltonian form.
	\item For a macroscopic many-body system, there exists a state of thermodynamic equilibrium, which, depending on external conditions, obeys axiomatically introduced corresponding distributions of statistical mechanics (microcanonical, canonical, or grand canonical ones). In terms of modern probability theory, these distributions are probability measures in the phase space of the system.
\end{enumerate}

Direct implementation of the Gibbs method comes down to solving two problems.

\begin{itemize}
	\item 
		The problem of describing interactions between the structural units of matter (atoms). It is generally assumed that this interaction can be represented in terms of instantaneous interatomic potentials, the very possibility of the existence of which is actually limited by the non-relativistic approximation, that is, the case of particles at rest. However, even in this approximation, systematic methods for finding interatomic potentials are unknown. Therefore, ``model potentials'' having the "correct" asymptotic behavior are usually used ~\cite{Kaplan, Kamberaj}.
	
	\item 
	The problem of exact calculation of the partition function for many-body systems with a given model interatomic potential. Despite enormous efforts over many years, the number of model interatomic potentials for which exact solutions have been found remains extremely limited~\cite{Baxter, Grosse, McCoy, Sutherland}. It is significant that there is still not a single exactly solvable three-dimensional model.

\end{itemize}

Besides the insurmountable \textbf{computational} difficulties, there are at least two unsolved \textbf{fundamental problems} in statistical mechanics.

\begin{enumerate}
	
	\item \textbf{Physical problem}. The zeroth law of thermodynamics, postulated in both phenomenological thermodynamics and statistical mechanics, is fundamentally incompatible with classical Newtonian mechanics~\cite{Uhlenbeck}. The essence of the matter is that within the framework of classical mechanics of a system of interacting particles, there is no mechanism leading to the irreversibility of the system's dynamics. Therefore, a consistent microscopic foundation of thermodynamics within the framework of classical mechanics is fundamentally impossible~\cite{Zakharov2025}.
	
	\textbf{Mathematical problem}. The concept of probability based on Lebesgue measure theory~\cite{Kolmogorov}, has an inevitable  mathematical ambiguity due to Ulam's theorem~\cite{Ulam1930,Ulam1960}, according to which there is no countably additive measure $\mu(A)$ defined for \textit{all} subsets $A$ of a set $E$ of power continuum~($\aleph_{1} $).
	In statistical mechanics, the set of elementary events~$E$ is the set of all points in the phase space~$\Gamma$ of the system~-- this is a set of power~($\aleph_{1} $). Therefore, different measures in the phase space of a system of particles correspond to different (generally speaking, \textbf{nonequivalent}) probability models~\cite{Khrennikov2009, Khrennikov2016, Khrennikov2023}. In particular, the question of the equivalence conditions of the basic probability measures (i.e., the microcanonical, canonical, and grand canonical ensembles) of statistical mechanics remains open~\cite{Ellis2000, Campa, Zhang2023}.
	
\end{enumerate}

One of the options for a microscopic explanation and justification of thermodynamics is the relativistic field model of a system of interacting particles (atoms). Within this model, the system consists of two subsystems, one of which are the atoms themselves, and the other of which is the field, which mediates the interactions between the atoms.

The fundamental difference between a relativistic field system of interacting particles and a non-relativistic one is that the set of degrees of freedom of the "atoms + field" system consists of a finite number of atomic degrees of freedom and an infinite set of field degrees of freedom. Essentially, the system of atoms is immersed in the variable field they create, which acts as a hidden, irremovable thermostat. Therefore, thermodynamic equilibrium in a system of interacting atoms is an equilibrium between the atoms, on one side, and the field (thermostat) they create, on the other. The energy of interatomic interactions is the energy of the field that mediates the interactions between the atoms. 

Thus, in non-relativistic dynamics in general and classical statistical mechanics in particular, the infinite number of field degrees of freedom of a system is completely ignored, and only a finite number of atomic degrees of freedom are considered. The evolution of the system as a whole depends on the initial conditions of both the atoms and the field. Therefore, ignoring the initial conditions for the field leads to unpredictability of both the dynamics of the system as a whole and the dynamics of the atoms in particular.

In turn, due to the finite propagation velocity of the field, the instantaneous configuration of the field generated by atoms depends on the history of the atomic dynamics. This leads to the phenomenon of heredity, including the effect of lag in interactions between atoms. The works ~\cite{Zakharov2016, Zakharov2019} have shown that lag in interatomic interactions is a real physical mechanism leading to the irreversibility of atomic dynamics. Thus, to explain the phenomenon of irreversibility from a microscopic perspective, there is no need to resort to probabilistic hypotheses.

\section{The concept of an auxiliary field. Lagrangian picture}

The concept of an auxiliary field was proposed and developed in the works~\cite{Zakharov2022, Zakharov2022-2, Zakharov2023}. By definition, an auxiliary field~$\varphi\left(\mathbf{r}, t\right)$ is a function that satisfies the following conditions:
\begin{enumerate}
	\item for atoms at rest~$\varphi\left(\mathbf{r}, t\right)$ is equivalent to a given central interatomic potential $v\left( r\right) $, $r =\left| \mathbf{r}\right| $;
	\item for atoms in motion~$\varphi\left(\mathbf{r}, t\right)$ is a covariant relativistic field. 
\end{enumerate}
For the class of interatomic potentials that can be represented as a Fourier integral
\begin{equation}\label{Four2}
	v\left( r\right) = \int\, \dfrac{d \mathbf{k} }{\left(2\pi \right)^{3} }\, \tilde{v}\left(k^{2} \right) \, e^{i \mathbf{k r}},
\end{equation}
the equation for the free auxiliary field has the form:
\begin{equation}\label{varphi-0,t}
	\left( \tilde{v}\left( -\square\right)\right)^{-1}  \varphi\left( \mathbf{r},t\right) = 0,
\end{equation}	
where $\square = \dfrac{\partial^{2}}{\partial x^{2}} + \dfrac{\partial^{2}}{\partial y^{2}} + \dfrac{\partial^{2}}{\partial z^{2}} - \dfrac{1}{c^{2}}\,\dfrac{\partial^{2}}{\partial t^{2}} $~is the d'Alembert operator.

To each singular point (pole) $k_{s}$ of the function~$\tilde{v}\left(k^{2} \right) $ on the complex plane of the variable~$k$ there corresponds an \textit{elementary auxiliary field}~$\varphi_{s} \left(\mathbf{r}, t \right) $ satisfying an equation of the Klein-Gordon type with, generally speaking, a complex parameter $\mu_{s}=i k_{s}$:
\begin{equation}\label{partial-s}
	\left( \square - \mu_{s}^{2}\right)^{\nu_{s}}\varphi_{s}\left(\mathbf{r}, t \right) = 0,
\end{equation} 
where~$\nu_{s}$~is the pole multiplicity.

The complete auxiliary field $\varphi\left(\mathbf{r}, t \right)$ is a superposition (linear combination) of the elementary auxiliary fields~$\varphi_{s}\left(\mathbf{r}, t \right)$
\begin{equation}\label{superpos}
	\varphi\left( \mathbf{r}, t\right) = \sum_{s} C_{s}\, \varphi_{s}\left( \mathbf{r}, t\right).
\end{equation}
Thus, within the framework of relativistic theory, the auxiliary field that ensures interatomic interactions is completely characterized by the set of singular points of the function $\tilde{v}\left( k^{2}\right) $, the existence of which is guaranteed by Liouville's theorem~\cite{Whittaker}.

We restrict ourselves to the case where the multiplicities of all poles of the function~$\tilde{v}\left( k^{2}\right) $ are equal to unity~$\nu_{s}=1$ and the parameters~$\mu_{s}$ are real.

The complete system of equations describing the dynamics of atoms and the auxiliary field they create follows from the variational principle for the action functional~\cite{Zakharov2022, Zakharov2022-2}:
\begin{equation}\label{Action}
	\begin{array}{r}
		{\displaystyle 	S=-\sum\limits_{a} m c\int \ ds_{a}-\sum_{s=1}^{n} \sum_{a}\frac{\gamma_{s}}{c}\int \varphi_{s}(x_{a}) \ ds_{a} }\\
		{\displaystyle + \sum_{s}\frac{\varkappa_{s}}{2c} \int   d^{4}x\,   \left( \partial_{\nu} \varphi_{s}(x)\, \partial^{\nu}\! \varphi_{s}(x) - \mu_{s}^2\varphi_{s}^2(x) \right),}
	\end{array}
\end{equation}
where $m$ and $s_{a}$~are the mass and world line of the $a$-th atom, respectively, $\gamma_{s}$ and $\varkappa_{s}$~are constants. 

Note that the action functional is based on the assumption that atoms do not undergo destruction over time, which in reality can only occur at sufficiently high speeds. Therefore, we will assume that the characteristic velocities of atoms, $v$, are much less than the speed of light, $c$:
\begin{equation}\label{v<<c}
	v \ll c.
\end{equation}
As a result, the actual expression for the action functional of the system ``atoms + auxiliary field'' has the form:
\begin{equation}\label{Action2}
	\begin{array}{r}
		{\displaystyle  	 S = \int  dt\left\lbrace \sum\limits_{a} \frac{m  \dot{\mathbf{R}}_{a}^{2}\left( t\right) }{2} -\sum_{s} \gamma_{s} \int\limits_{\left( V\right) } d\mathbf{r} \left( \sum_{a} \delta \left(\mathbf{r - R}_{a}\left(t \right)  \right) \right)  \varphi_{s}(\mathbf{r}, t)  \right.  }\\
		{\displaystyle \left.   + \sum_{s}\frac{\varkappa_{s}}{2} \int\limits_{\left( V\right) }  d\mathbf{r} \, \left[   \left(  \frac{\partial \varphi_{s}(\mathbf{r}, t)}{c\, \partial t}\right)^{2}   -  \left(\nabla \varphi_{s}(\mathbf{r}, t) \right)^{2} -  \mu_{s}^2\varphi_{s}^2(\mathbf{r}, t)\right] \right\rbrace ,  }
	\end{array}
\end{equation}
%
%
where $V$~is the volume of the system.

The integrand in this formula is the Lagrangian  
\begin{equation}\label{Lagrangian}
L \left(\mathbf{R}_{a}\left( t\right),   \dot{\mathbf{R}}_{a}\left( t\right); \left\lbrace  \varphi_{s}(\mathbf{r}, t) \right\rbrace,  \left\lbrace  \dot {\varphi}_{s}(\mathbf{r}, t) \right\rbrace, \left\lbrace  \nabla \varphi_{s}(\mathbf{r}, t) \right\rbrace \right)
\end{equation}
of a system consisting of atoms and the composite auxiliary field they create.
From here follow both the equations of the dynamics of atoms 
\begin{equation}\label{atoms}
	m \ddot{\mathbf{R}}_{a}\left( t\right) + \sum_{s}\, \gamma_{s}\nabla \varphi_{s}\left(\mathbf{R}_{a}\left( t\right)  \right) = 0,
\end{equation}
and the equations of the evolution of the elementary auxiliary fields of the system:
\begin{equation}\label{el-field}
	\left(\square -\mu_{s}^{2} \right) \varphi_{s}\left( \mathbf{r}, t\right) = \frac{\gamma_{s}}{\varkappa_{s}} \sum_{a} \delta\left( \mathbf{r - R}_{a}\left( t\right)  \right).
\end{equation}
The general solution of the field dynamics equation
$\varphi_{s}\left( \mathbf{r}, t\right)$ can be found, for example, using the well-known method of parameter variation and represented as the Duhamel integral~\cite{Courant}.

It is essential that this solution depends on the trajectories~$\mathbf{R}_{a}\left( t'\right) $ of all atoms ($t'\leq t$), i.e. on the entire history of the system of atoms up to the moment of time~$t$. 

Substitution solutions of the equations~\eqref{el-field} into the equations~\eqref{atoms} yields to a closed system of functional-differential equations with respect to~$\mathbf{R}_{a}\left( t\right) $.
It should be noted that the theory of such equations is still in its early stages of development~\cite{Berezansky,Xu}.
In particular, even simple functional-differential equations with a single unknown function are infinite-dimensional in nature and possess an incomparably richer structure than their ordinary differential equation counterparts~\cite{Baker, Corduneanu}.

Thus, eliminating field variables in the dynamics of interacting atoms is, in principle, possible. However, solving, or even qualitatively analyzing, the problems that arise in this way are beyond the capabilities of modern mathematics.

In this regard, let us consider the dynamics of a system of atoms interacting through an auxiliary field within the framework of the Hamiltonian picture.

\vskip2cm

\section{Hamiltonian of a system consisting of atoms and an auxiliary field}

Let us pass from the elementary auxiliary fields $\varphi_{s}\left(\mathbf{r}, t \right) $ to their Fourier components $\tilde{v}_{s}\left(\mathbf{k}, t \right)$~\cite{Haar}:
\begin{equation}\label{Four-0}
	\begin{array}{r}
		{\displaystyle 	\varphi_{s}\left( \mathbf{r}, t\right) = \sum_{\mathbf{k}} \tilde{\varphi}_{s}\left(\mathbf{k}, t \right) e^{i \mathbf{k r}}, }\\
		{\displaystyle \tilde{\varphi}_{s}\left(\mathbf{k}, t \right) = \frac{1}{V}\int\limits_{\left( V\right) } \varphi_{s}\left(\mathbf{r}, t \right) e^{-i \mathbf{k r}}\, d\mathbf{r}. }
	\end{array}
\end{equation}
Note that the Fourier components~$\varphi_{s} \left(\mathbf{k}, t \right) $ are complex-valued and have the following properties
\begin{equation}\label{var-phi}
	\tilde{\varphi}_{s}\left(-\mathbf{k}, t \right) = \tilde{\varphi}^{*}_{s}\left(\mathbf{k}, t \right).
\end{equation}
To each complex-valued function~$\varphi_{s}\left(\mathbf{k}, t \right) $ we assign a pair of real field variables  ${\psi}_{s}\left(\mathbf{k}, t \right)$ and ${\chi}_{s}\left(\mathbf{k}, t \right)$:
\begin{equation}\label{var--phi}
	\tilde{\varphi}_{s}\left(\pm \mathbf{k}, t \right) = {\psi}_{s}\left(\mathbf{k}, t \right) \pm i{\chi}_{s}\left(\mathbf{k}, t \right),
\end{equation}
that have the following properties
\begin{equation}\label{pm}
	\left\lbrace 
	\begin{array}{l}
		{\displaystyle {\psi}_{s}\left(-\mathbf{k}, t \right) = {\psi}_{s}\left(\mathbf{k}, t \right),}\\
		{\displaystyle {\chi}_{s}\left(-\mathbf{k}, t \right) = -{\chi}_{s}\left(\mathbf{k}, t \right)}.
	\end{array}
	\right. 
\end{equation}

Let's transform the Lagrangian $L \left(\mathbf{R}_{a}\left( t\right),   \dot{\mathbf{R}}_{a}\left( t\right); \left\lbrace  \varphi_{s}(\mathbf{r}, t) \right\rbrace,  \left\lbrace  \dot {\varphi}_{s}(\mathbf{r}, t) \right\rbrace, \left\lbrace  \nabla \varphi_{s}(\mathbf{r}, t) \right\rbrace \right)$ to new variables~$\mathbf{R}_{a}\left( t\right),   \dot{\mathbf{R}}_{a}\left( t\right)$;    $ \psi_{s}\left(\mathbf{k}, t \right), \dot{\psi}_{s}\left(\mathbf{k}, t \right)$;  $  \chi_{s}\left(\mathbf{k}, t \right), \dot{\chi}_{s}\left(\mathbf{k}, t \right)  $: 
\begin{equation}\label{Lagrange1}
	\begin{array}{l}
		{\displaystyle 	L \left(\mathbf{R}_{a}\left( t\right),   \dot{\mathbf{R}}_{a}\left( t\right);   {\psi}_{s}(\mathbf{k}, t),   \dot{\psi}_{s}\left(\mathbf{k}, t\right); {\chi}_{s}(\mathbf{k}, t),   \dot{\chi}_{s}\left(\mathbf{k}, t\right) \right)  =  \sum\limits_{a} \frac{m \dot{\mathbf{R}}_{a}^{2}\left( t\right) }{2}}\\
		{\displaystyle  - \sum_{s} \gamma_{s} \sum_{a, \mathbf{k}}\,\left[  \psi_{s} \left(\mathbf{k}, t \right)\, \cos\left(\mathbf{k R}_{a}\left(t \right)  \right)  -   \chi_{s}\left( \mathbf{k}, t\right) \sin\left( \mathbf{k R}_{a}\left(t \right) \right)   \right]  }\\ 
		{\displaystyle  + \sum_{s}\frac{V\varkappa_{s}}{2c^{2}} 
			\sum_{\mathbf{k}} \left[ \dot{\psi}_{s}^{2}\left(\mathbf{k}, t\right)  - c^{2}\left(k^{2} + \mu^{2}_{s} \right) {\psi}_{s}^{2}\left(\mathbf{k}, t \right) \right] }\\
		{\displaystyle  + \sum_{s}\frac{V\varkappa_{s}}{2c^{2}} 
			\sum_{\mathbf{k}} \left[ \dot{\chi}_{s}^{2}\left(\mathbf{k}, t\right)  - c^{2}\left(k^{2} + \mu^{2}_{s} \right) {\chi}_{s}^{2}\left(\mathbf{k}, t \right) \right] .}\\
	\end{array}
\end{equation}

We define the momenta of the atoms $\mathbf{P}_{a}\left( t \right) $ and of the elementary auxiliary fields $p_{s}\left( \mathbf{k}, t\right),\ \mathfrak{p}_{s} \left( \mathbf{k}, t\right) $ by the relations:
\begin{equation}\label{p_a}
	\left\lbrace 
	\begin{array}{l}
		{\displaystyle \mathbf{P}_{a} \left( t \right) = \frac{ \partial L}{\partial \dot{\mathbf{R}}_{a} \left( t \right) } = m \dot{\mathbf{R}}_{a}\left( t\right); }\\ 
		{\displaystyle p_{s}\left( \mathbf{k}, t\right) = \frac{\partial L}{\partial  \dot{\psi}_{s}\left(\mathbf{k}, t \right)}  = \frac{V\varkappa_{s}}{c^{2}} \, \dot{\psi}_{s} \left(\mathbf{k},t \right) ;}\\
		{\displaystyle \mathfrak{p}_{s}\left( \mathbf{k}, t\right) = \frac{\partial L}{\partial  \dot{\chi}_{s}\left(\mathbf{k}, t \right)}  = \frac{V\varkappa_{s}}{c^{2}} \, \dot{\chi}_{s} \left(\mathbf{k},t \right) .}
	\end{array}
	\right. 
\end{equation}

By performing the Legendre transformation, we find the Hamiltonian of a system consisting of atoms and auxiliary fields

\begin{equation}\label{Hamilton1}
	\begin{array}{r}
		{\displaystyle 	H \left(\mathbf{R}_{a}\left( t\right), {\mathbf{P}}_{a}\left( t\right);   {\psi}_{s}(\mathbf{k}, t),  {p}_{s}(\mathbf{k}, t) ; \chi_{s}\left(\mathbf{k}, t \right),  \mathfrak{p}_{s}\left(\mathbf{k}, t \right) \right)  = \sum\limits_{a} \frac{\mathbf{P}_{a}^{2}\left( t\right) }{2 m}  } \\
		{\displaystyle  + 
			\sum_{s} \gamma_{s} \sum_{a, \mathbf{k}}\,\left[  \psi_{s} \left(\mathbf{k}, t \right)\, \cos\left(\mathbf{k R}_{a}\left(t \right)  \right)   - \chi_{s}\left( \mathbf{k}, t\right) \sin\left( \mathbf{k R}_{a}\left(t \right) \right)   \right]} \\
			{\displaystyle + \sum_{s} \sum_{\mathbf{k}} \left[   \frac{c^{2}}{V\varkappa_{s}} \frac{p_{s}^{2} \left( \mathbf{k}, t\right)}{2}  +   V\varkappa_{s} \left(k^{2} + \mu^{2}_{s} \right) \frac{{\psi}_{s}^{2}\left(\mathbf{k}, t \right)}{2}  \right] }\\
		{\displaystyle + \sum_{s} \sum_{\mathbf{k}} \left[   \frac{c^{2}}{V\varkappa_{s}} \frac{\mathfrak{p}_{s}^{2} \left( \mathbf{k}, t\right)}{2}  +   {V\varkappa_{s}} \left(k^{2} + \mu^{2}_{s} \right) \frac{{\chi}_{s}^{2}\left(\mathbf{k}, t \right)}{2}  \right] .}
	\end{array}
\end{equation}

The first term on the right-hand side of this formula is the Hamiltonian of the free atoms subsystem; the third and fourth terms are the Hamiltonians of the free massive (due to the parameters $\mu_{s}$) oscillators subsystems with coordinates ${\psi}_{s}\left(\mathbf{k}, t \right)$, ${\chi}_{s}\left(\mathbf{k}, t \right)$ and momenta $p_{s} \left(\mathbf{k}, t\right)$, $\mathfrak{p}_{s} \left(\mathbf{k}, t\right)$, respec\-tively. The second term contains the interaction between the atoms on one side and the auxiliary field on the other. It ensures the exchange of energy between these two subsystems.

We will interpret thermodynamic equilibrium in the system as a whole as equilibrium between these subsystems.

It should be noted that the exact relativistic Hamiltonian of a system of interacting atoms is fundamentally different from the approximate non-relativistic Hamiltonian
\begin{equation}\label{non-rel-Ham}
	H_{\mathrm{nr}}\left( \mathbf{R}_{a}\left( t\right), {\mathbf{P}}_{a}\left( t\right)\right)  = \frac{\mathbf{P}_{a}^{2}\left( t\right) }{2 m} + \sum\limits_{a<a'}v\left( \mathbf{R}_{a} \left( t\right) -  \mathbf{R}_{a'} \left( t\right) \right) 
\end{equation}
of the same system. 
Namely, the interaction energy between atoms in the relativistic Hamiltonian is the sum of one-particle terms, whereas in the non-relativistic Hamiltonian the interaction energy is the sum of two-particle terms.
Therefore, the complex problem of decoupling of atomic coordinates, which arises when calculating the configuration integral in the non-relativistic approximation 
\begin{equation}\label{config}
	Z_{\mathrm{conf}} = \idotsint\limits_{\left( V\right) } \left( \prod\limits_{a=1}^{N} d\mathbf{R}_{a}\right) \ \exp\left\lbrace -\beta \sum\limits_{a<a'}v\left( \mathbf{R}_{a} -  \mathbf{R}_{a'}  \right) \right\rbrace, 
\end{equation}
does not arise at all within the relativistic field theory.

\section{Canonical partition function}
The partition function of the system within the relativistic model has the form

\begin{equation}\label{part-fun}
	\begin{array}{r}
		{\displaystyle  Z = \frac{1}{N!} \idotsint \left( \prod_{s, \mathbf{k}} d\psi_{s}\left(\mathbf{k} \right)  dp_{s}\left(\mathbf{k} \right)  d\chi_{s}\left(\mathbf{k} \right)  d\mathfrak{p}_{s}\left(\mathbf{k} \right)  \right) \left( \prod_{a}d\mathbf{R}_{a}\, d\mathbf{P}_{a} \right) }\\
		{\displaystyle \exp \left[ -\beta H\left(\mathbf{R}_{a}, {\mathbf{P}}_{a};  {\psi}_{s}(\mathbf{k}),  {p}_{s}(\mathbf{k}); \chi_{s}\left(\mathbf{k}\right),  \mathfrak{p}_{s}\left(\mathbf{k} \right) \right)  \right],  }
	\end{array}
\end{equation}
where $H\left(\mathbf{R}_{a}, {\mathbf{P}}_{a};   {\psi}_{s}(\mathbf{k}),  {p}_{s}(\mathbf{k}) ; \chi_{s}\left(\mathbf{k}\right),  \mathfrak{p}_{s}\left(\mathbf{k} \right) \right)$ is the Hamiltonian~\eqref{Hamilton1}, $\beta = 1/T$.

Let us transform this multiple (essentially infinite-dimensional) integral ~\eqref{part-fun} into an iterative integral. 



\subsection{Integration over the atomic variables~$\mathbf{R}_{a}, \mathbf{P}_{a}$.}
Integration over atomic momenta~$\mathbf{P}_{a}$ is quite trivial:
\begin{equation}\label{P-a}
	\idotsint \left(\prod_{a} d\mathbf{P}_{a}\right) \exp{\left[ -\beta\sum_{a}\frac{\mathbf{P}_{a}^{2}}{2m}\right] } = \left(\frac{2\pi m}{\beta} \right)^{\frac{3N}{2}}. 
\end{equation}

The integral over atomic coordinates~$\mathbf{R}_{a}$ has the form 
\begin{equation}\label{R-a}
	\begin{array}{c}
	{\displaystyle V^{N}\,\idotsint\limits_{\left( V\otimes V\otimes \cdots \otimes V\right) } \left(\prod_{a}  \frac{d\mathbf{R}_{a}}{V}   \exp\left[-\beta\sum_{a, s, \mathbf{k}}\gamma_{s} \left\lbrace \psi_{s} \left(\mathbf{k} \right)\, \cos\left(\mathbf{k R}_{a}  \right)  -  \chi_{s} \left(\mathbf{k} \right)\, \sin\left(\mathbf{k R}_{a}  \right)   \right\rbrace  \right]\right)}\\
	{\displaystyle = V^{N} \left[ Y\left(\beta  \gamma_{s}\sqrt{\psi_{s}^{2} \left(\mathbf{k} \right) + \chi_{s}^{2} \left(\mathbf{k} \right) }  \right) \right] ^{N}, }
	\end{array}
\end{equation}
where 
\begin{equation}\label{Y-1}
\begin{array}{c}
{\displaystyle 	Y\left(\beta  \gamma_{s}\sqrt{\psi_{s}^{2} \left(\mathbf{k} \right) + \chi_{s}^{2} \left(\mathbf{k} \right) }  \right) }\\
{\displaystyle  =	\int\limits_{\left( V\right) } \frac{d\mathbf{R}}{V}  \exp\left[-\beta\sum_{s, \mathbf{k}}\gamma_{s}   \sqrt{\psi_{s}^{2} \left(\mathbf{k} \right) + \chi_{s}^{2} \left(\mathbf{k} \right) } \cos\left(2\pi Z_{s}\left( \mathbf{k}, \mathbf{R}\right)   \right)   \right], }
\end{array}
\end{equation}
and $Z_{s} \left(\mathbf{k}, \mathbf{R} \right) = \left(\frac{\mathbf{kR} + \alpha_{s}\left(\mathbf{k}\right) }  {2\pi} \right) $.

\begin{equation}\label{psi-chi}
\begin{array}{c}
{\displaystyle   Z = \frac{V^{N}}{N!} \idotsint \left( \prod_{s, \mathbf{k}} d\psi_{s}\left(\mathbf{k} \right)  dp_{s}\left(\mathbf{k} \right)  d\chi_{s}\left(\mathbf{k} \right)  d\mathfrak{p}_{s}\left(\mathbf{k} \right)  \right)    }\\
{\displaystyle \times \exp{\left[ - \beta H_{\mathrm{field}} \left(\beta;  {\psi}_{s}(\mathbf{k}),  {p}_{s}(\mathbf{k}) ; \chi_{s}\left(\mathbf{k} \right),  \mathfrak{p}_{s}\left(\mathbf{k}\right) \right) \right] }  }\\
\end{array}
\end{equation}
where 
\begin{equation}\label{HamField}
	\begin{array}{c}
		{\displaystyle 	H_{\mathrm{field}} \left( \beta; {\psi}_{s}(\mathbf{k}),  {p}_{s}(\mathbf{k}) ; \chi_{s}\left(\mathbf{k} \right),  \mathfrak{p}_{s}\left(\mathbf{k}\right) \right)  } \\
		{\displaystyle =\sum_{s} \sum_{\mathbf{k}} \left[   \frac{c^{2}}{V\varkappa_{s}} \frac{p_{s}^{2} \left( \mathbf{k}\right)}{2}  +   V\varkappa_{s} \left(k^{2} + \mu^{2}_{s} \right) \frac{{\psi}_{s}^{2}\left(\mathbf{k} \right)}{2}  \right] }\\
		{\displaystyle + \sum_{s} \sum_{\mathbf{k}}  \left[   \frac{c^{2}}{V\varkappa_{s}} \frac{\mathfrak{p}_{s}^{2} \left( \mathbf{k}\right)}{2}  +   {V\varkappa_{s}} \left(k^{2} + \mu^{2}_{s} \right) \frac{{\chi}_{s}^{2}\left(\mathbf{k} \right)}{2}  \right] }\\
			{\displaystyle - \frac{N}{\beta} \ln Y\left(\beta  \gamma_{s}\sqrt{\psi_{s}^{2} \left(\mathbf{k} \right) + \chi_{s}^{2} \left(\mathbf{k} \right) }  \right)  }.
	\end{array}
\end{equation}
The first and second terms on the right-hand side of this formula constitute the Hamiltonian of the free auxiliary field $\varphi\left( \mathbf{r}, t\right) $, and the last term is an exact description of the self-action of the auxiliary field through all the atoms of the system.


Let's rearrange the terms on the right side of formula~\eqref{HamField}:
\begin{equation}\label{HamField2}
	\begin{array}{c}
		{\displaystyle 	H_{\mathrm{field}} \left( \beta; {\psi}_{s}(\mathbf{k}),  {p}_{s}(\mathbf{k}) ; \chi_{s}\left(\mathbf{k} \right),  \mathfrak{p}_{s}\left(\mathbf{k}\right) \right)  = 
			 \sum_{s} \sum_{\mathbf{k}} \Biggl\{    \frac{c^{2}}{V\varkappa_{s}}\left(  \frac{p_{s}^{2} \left( \mathbf{k}\right)}{2}  + \frac{\mathfrak{p}_{s}^{2} \left( \mathbf{k}\right)}{2} \right)   }\\
		{\displaystyle + \left[     {V\varkappa_{s}} \left(k^{2} + \mu^{2}_{s} \right) \left( \frac{{\psi}_{s}^{2}\left(\mathbf{k} \right)}{2} +  \frac{{\chi}_{s}^{2}\left(\mathbf{k} \right)}{2} \right)\right] \Biggr\}   - \frac{N}{\beta} \ln Y\left(\beta  \gamma_{s}\sqrt{\psi_{s}^{2} \left(\mathbf{k} \right) + \chi_{s}^{2} \left(\mathbf{k} \right) }  \right)   . }
		\end{array}
\end{equation}

\subsection{Weyl's theorem}
To calculate the integral~\eqref{Y-1}, we use Weyl's theorem~\cite{Weyl, Polya}.\\
\textit{Theorem}. Suppose
\begin{enumerate}
	\item  $x_{\nu}\left( t\right)  = a_{\nu}\, + b_{\nu} t \left( \nu = 1, 2, \ldots, N\right)  $, where $a_{\nu}$, $b_{\nu}$ are constants; 
	\item the coefficients $b_{\nu}$ are rationally independent  numbers (i.e. from the relation $\sum_{\nu}n_{\nu} b_{\nu}=0$ with rational coefficients $n_{\nu}$ should follow $n_{1}=n_{2}=\ldots=n_{N}=0$);
	\item the function $f\left(x_{1}, x_{2}, \ldots, x_{N} \right)$ is periodic in all variables $x_{\nu}$ with period equal to $1$ and integrable in the
	cube $0\leq x_{\nu} \leq 1$. 
\end{enumerate}
Then
\begin{equation}\label{Weyl-}
\begin{array}{r}
{\displaystyle \lim\limits_{t\to\infty}\frac{1}{t}\int\limits_{0}^{\infty}f\left(x_{1}\left( t\right), x_{2}\left( t\right),\ldots, x_{N}\left( t\right) \right)\, dt  }\\
{\displaystyle = \int\limits_{0}^{1}\int\limits_{0}^{1}\cdots \int\limits_{0}^{1}f\left(x_{1}, x_{2}, \ldots x_{N} \right) \left(\prod_{s=1}^{N}dx_{s} \right)  }.	
\end{array}	 
\eqref{HamField}\end{equation}
This theorem has an obvious generalization to the case of a three-dimensional (vector) variable $\mathbf{R}$ instead of a one-dimensional one $t$.

The integrand of this integral satisfies the Weyl conditions:
\begin{enumerate}
	\item it is periodic in all variables $Z_{s} \left(\mathbf{k}, \mathbf{R} \right) $ with period of~1;
	\item variables $Z_{s}\left( \mathbf{k, R}\right) $ are linear functions of the integration variable $\mathbf{R}$;
	\item the condition of rational independence of the variables $\mathbf{k}$ holds almost everywhere, i.e., up to subsets of zero Lebesgue measure (a detailed analysis of this condition is carried out in the work~\cite{Zakharov1990}).
\end{enumerate}

The interpretation of Weyl's theorem is as follows. Rational independence of quantities implies their incommensurability. The set of mutually commensurable quantities is countable, while the set of all mutually incommensurable quantities has the cardinality of the continuum. Therefore, the Lebesgue measure of rationally dependent quantities is zero.

The application of the Weyl's theorem reduces the integral over the variable $\mathbf{R}$ to a multiple integral over the variables $z_{s}\left(\mathbf{k, R}\right) $, which is a product of single integrals:
\begin{equation}\label{Weyl}
	\begin{array}{r}
{\displaystyle 	Y\left(\beta  \gamma_{s}\sqrt{\psi_{s}^{2} \left(\mathbf{k} \right) + \chi_{s}^{2} \left(\mathbf{k} \right) }  \right) = \prod_{s, \mathbf{k}}\int\limits_{0}^{1}  dz_{s}\left( \mathbf{k, R}\right) }\\
{\displaystyle \times \exp{\left[ -\beta \gamma_{s}   \sqrt{\psi_{s}^{2} \left(\mathbf{k} \right) + \chi_{s}^{2} \left(\mathbf{k} \right) } \cos\left(2\pi Z_{s}\left( \mathbf{k, R}\right)   \right)  \right]  }  }\\
{\displaystyle 
		= \prod_{s, \mathbf{k}}  I_{0} \left(\beta \gamma_{s}   \sqrt{\psi_{s}^{2} \left(\mathbf{k} \right) + \chi_{s}^{2} \left(\mathbf{k} \right) } \right),}
\end{array}
\end{equation}
where $I_{0}(x)$~is the modified Bessel function.

Substituting this expression for into~\eqref{HamField2}, we obtain
\begin{equation}\label{Ham-2}
	\begin{array}{r}
		{\displaystyle H_{\mathrm{field}}\left(\beta; \psi_{s} \left(\mathbf{k} \right), \chi_{s} \left(\mathbf{k} \right); p_{s} \left(\mathbf{k} \right),  \mathfrak{p}_{s} \left(\mathbf{k} \right) \right) = \sum_{s,\mathbf{k}}\Biggl[ \frac{c^{2}}{V\varkappa_{s}}\left[ \frac{p_{s}^{2} \left( \mathbf{k}\right)}{2} +  \frac{\mathfrak{p}_{s}^{2} \left( \mathbf{k} \right)}{2}  \right] }\\
		{\displaystyle +  \frac{N}{\beta} \left(  \frac{\beta\varkappa_{s} \left(k^{2} + \mu^{2}_{s} \right)}{2n}     \left[      {{\psi}_{s}^{2}\left(\mathbf{k} \right)}  +   { {\chi}_{s}^{2}\left(\mathbf{k} \right)}\right] -  \ln I_{0} \left(\beta \gamma_{s}   \sqrt{\psi_{s}^{2} \left(\mathbf{k} \right) + \chi_{s}^{2} \left(\mathbf{k} \right) } \right) \right) \Biggr]}, 
	\end{array}
\end{equation}
where $n=N/V$.

Thus, the problem of calculating the canonical partition function of a system of atoms interacting through a covariant auxiliary field is reduced to calculating the partition function of a system of independent nonlinear oscillators.

In other words, the relativistic statistical thermodynamics of a system of interacting atoms is equivalent to the statistical thermodynamics of an ideal gas of quasiparticles, whose Hamiltonian has the form
\begin{equation}\label{H-quasi}
	\begin{array}{r}
		{\displaystyle H_{\mathrm{quasi}}\left(\beta; \psi_{s} \left(\mathbf{k} \right), \chi_{s} \left(\mathbf{k} \right); p_{s} \left(\mathbf{k} \right),  \mathfrak{p}_{s} \left(\mathbf{k} \right) \right) =  \frac{c^{2}}{V\varkappa_{s}}\left[ \frac{p_{s}^{2} \left( \mathbf{k}\right)}{2} +  \frac{\mathfrak{p}_{s}^{2} \left( \mathbf{k} \right)}{2}  \right] }\\
		{\displaystyle +  \frac{N}{\beta} \left(  \frac{\beta\varkappa_{s} \left(k^{2} + \mu^{2}_{s} \right)}{2n}     \left[      {{\psi}_{s}^{2}\left(\mathbf{k} \right)}  +   { {\chi}_{s}^{2}\left(\mathbf{k} \right)}\right] -  \ln I_{0} \left(\beta \gamma_{s}   \sqrt{\psi_{s}^{2} \left(\mathbf{k} \right) + \chi_{s}^{2} \left(\mathbf{k} \right) } \right) \right) }. 
	\end{array}
\end{equation}
Note that each of these quasiparticles has two degrees of freedom~$\psi_{s}\left(k\right)$, $\chi_{s}\left(k\right)$. The set of all these quasiparticles forms a nonlinear composite effective auxiliary field~$\Phi\left(\mathbf{r}, t \right) $, whose Hamiltonian is defined by expression~\eqref{Ham-2}. 

The specific features of this field.
\begin{enumerate}
	\item It includes both a free auxiliary field and an additional term that takes into account the interaction between the atoms and the auxiliary field.
	\item It depends on the system temperature $T=1/\beta $.
	\item Although it is essentially nonlinear, it still represents a superposition of independent elementary (nonlinear) components.
\end{enumerate}

\subsection{Partition functions of quasiparticles}
The multiple integral over field variables has the form:
\begin{equation}\label{ps(k)}
	\begin{array}{r}
		{\displaystyle Z_{f} = \idotsint \left( \prod_{s,\mathbf{k}} d\psi_{s}\left( \mathbf{k}\right)\, dp_{s}\left( \mathbf{k}\right)\,d\chi_{s}\left( \mathbf{k}\right)  d\mathfrak{p}_{s}\left( \mathbf{k}\right) \right)}\\
		{\displaystyle \times  
			\exp\left\lbrace -\beta H_{\mathrm{field}}\left(\psi_{s} \left(\mathbf{k} \right), \chi_{s} \left(\mathbf{k} \right); p_{s} \left(\mathbf{k} \right),  \mathfrak{p}_{s} \left(\mathbf{k} \right); \beta \right)   \right\rbrace  },
	\end{array}
\end{equation}
where    

It should be noted that formula~\eqref{ps(k)} is the partition function of the composite nonlinear effective  field~$\Phi\left(\mathbf{r}, t \right) $, whose Hamiltonian is defined by expression~\eqref{Ham-2}.

Beside, the very multiple integral~\eqref{ps(k)} splits into a product of integrals of much lesser multiplicity, each of which corresponds to certain values of~$s$ and~$\mathbf{k}$:
\begin{equation}\label{dp-dpsi}
\begin{array}{r}
{\displaystyle Z_{s}\left( \mathbf{k}\right) = \iint\limits_{-\infty}^{\infty}  dp_{s}\left( \mathbf{k}\right) d\mathfrak{p}_{s}\left( \mathbf{k}\right) \exp \left[-\beta \frac{c^{2}}{V\varkappa_{s}}\left( \frac{p_{s}^{2} \left( \mathbf{k}\right)}{2} +  \frac{\mathfrak{p}_{s}^{2} \left( \mathbf{k} \right)}{2}  \right) \right]}\\
{\displaystyle \times \iint\limits_{-\infty}^{\infty} d\psi_{s}\left( \mathbf{k}\right) d\chi_{s}\left( \mathbf{k}\right) \exp \Biggl[-\beta N \Biggl\{  \frac{\varkappa_{s} \left(k^{2} + \mu^{2}_{s} \right)}{2n}     \left[      {{\psi}_{s}^{2}\left(\mathbf{k} \right)}  +   { {\chi}_{s}^{2}\left(\mathbf{k} \right)}\right] }\\ 
{\displaystyle - \frac{1}{\beta} \ln I_{0} \left(\beta \gamma_{s}   \sqrt{\psi_{s}^{2} \left(\mathbf{k} \right) + \chi_{s}^{2} \left(\mathbf{k} \right) } \right)  \Biggr\}   \Biggr]},
\end{array}
\end{equation}

The integral over variables $p_{s}\left(\mathbf{k} \right) $ and $\mathfrak{p}_{s}\left(\mathbf{k} \right)$ is calculated simply.  

To investigate the double integral over the variables $\psi_{s}\left(\mathbf{k} \right)$ and $\chi_{s}\left(\mathbf{k} \right)$, we use the asymptotic behavior of function~$\ln \left(I_{0}\left( \beta \gamma_{s}   \sqrt{\psi_{s}^{2} \left(\mathbf{k} \right) + \chi_{s}^{2} \left(\mathbf{k} \right) } \right) \right)$  for both small and large values of the variable~$x = \beta \gamma_{s}   \sqrt{\psi_{s}^{2} \left(\mathbf{k} \right) + \chi_{s}^{2} \left(\mathbf{k} \right) }$:
\begin{equation}\label{ln-I0}
	\ln \left(I_{0}\left( x\right)\right) \approx \left\lbrace \begin{array}{l}
		{\displaystyle \frac{x^{2}}{4} -\frac{x^{4}}{64} +\ldots, \ \ \ \ \ \ \ \text{if}\ x \lesssim 1;}\\ \\ 
		{\displaystyle x - \frac{1}{2}\ln\left(2\pi x \right)+ \ldots,  \text{if}\ x \gg 1.}
	\end{array}
	\right. 
\end{equation}
The convergence of this integral takes place under the condition
\begin{equation}\label{thermo}
\left\lbrace \frac{\varkappa_{s}\left(k^{2} + \mu_{s}^{2}\right) }{2n\beta \gamma_{s}^{2}}\right\rbrace   >  \frac{1}{4}.
\end{equation}
Since this condition must be satisfied for all auxiliary fields~$\varphi_{s}\left(\mathbf{r}, t \right) $ of the system and for all values of~$ \mathbf{k}$, then we have
\begin{equation}\label{thermo2}
	\min_{s, k}\left\lbrace \frac{\varkappa_{s}\left(k^{2} + \mu_{s}^{2}\right) }{2n\beta \gamma_{s}^{2}}\right\rbrace  = \min_{s} \left\lbrace \frac{\varkappa_{s}\mu_{s}^{2}}{2n\beta\gamma_{s}^{2}}\right\rbrace  >  \frac{1}{4}.
\end{equation}
We will assume that condition~\eqref{thermo} is satisfied. 

Note that condition~\eqref{thermo2} essentially represents a limitation on the system temperature
\begin{equation}\label{T-crit}
	T > T_{\mathrm{crit}} = \min_{s} \left\lbrace \frac{n \gamma_{s}^{2}}{2\varkappa_{s}\mu_{s}^{2}} \right\rbrace. 
\end{equation}
Thus, within the framework of this theory there exists a singular temperature~$T_{\mathrm{crit}}$, which depends on the particle number density~$n$ and the parameters~$\varkappa_{s},\ \mu_{s}, \ \gamma_{s}$ of the auxiliary fields.
$T_{\mathrm{crit}}$ can be interpreted as a phase transition point.

\subsubsection{Quadratic approximation for the Hamiltonian of the effective elementary auxiliary field}
Restricting ourselves to the first non-vanishing term in the expansion of the Hamiltonian~\eqref{Ham-2} for the $s$-th effective elementary field, we obtain
\begin{equation}\label{H(s)}
	\begin{array}{r}
		{\displaystyle H_{s} = \sum_{\mathbf{k}} \Biggl[\frac{c^{2}}{V\varkappa_{s}}\left[ \frac{p_{s}^{2} \left( \mathbf{k}\right)}{2} +  \frac{\mathfrak{p}_{s}^{2} \left( \mathbf{k} \right)}{2}  \right]  }\\
		{\displaystyle + N \left( \frac{\varkappa_{s}\left(k^{2} + \mu_{s}^{2} \right)  }{2n} - \frac{\beta \gamma_{s}^{2}}{4} \right)     \left[      {{\psi}_{s}^{2}\left(\mathbf{k} \right)}  +   { {\chi}_{s}^{2}\left(\mathbf{k} \right)}\right] \Biggr] .}
	\end{array}
\end{equation}

Its partition function~$Z_{s}\left(k \right) $ is
\begin{equation}
Z_{s}\left( \mathbf{k}\right) = \left(\frac{2\pi}{\beta \omega_{s}\left(\mathbf{k}, \beta \right)  } \right)^{2}, 
\end{equation}
where 
\begin{equation}\label{dispers}
	\omega_{s}\left( \mathbf{k}, \beta\right) = c\sqrt{k^{2}+\left( \mu_{s}^{2} -\frac{n}{2\varkappa_{s}}\beta\gamma_{s}^{2}\right) }
\end{equation}
is the dispersion law of oscillators of $s$-th elementary effective field~$\Phi\left(\mathbf{r}, t; \beta \right) $ in the quadratic approximation for the Hamiltonian.
From the non-negativity of the expression under the root for all values of $k$, a limitation on the temperature of the system follows
\begin{equation}\label{lim-T}
T = \beta^{-1}\geq \max_{s} \left\lbrace \frac{n \gamma_{s}^{2}}{2\varkappa_{s} \mu_{s}^{2}}\right\rbrace. 
\end{equation}

We also note that the interaction between atoms and elementary auxiliary fields~$\varphi_{s}\left(\mathbf{r},t \right)$ can be taken into account by renormalizion the mass parameters~$\mu_{s}$ of these fields:
\begin{equation}\label{mu-s}
	\mu_{s} \Rightarrow \sqrt{ \mu_{s}^{2} -\frac{n}{2\varkappa_{s}}\beta\gamma_{s}^{2}}.
\end{equation}

Thus, the expression for the classical partition function of a system of atoms interacting via a covariant composite auxiliary field, by the condition~\eqref{lim-T}, has the following form:
\begin{equation}\label{Z--full}
Z\left(N, V, T \right)  = \frac{V^{N}}{N!}\left(\frac{2\pi m}{\beta} \right)^{\frac{3N}{2}} \left[   \prod_{s,\mathbf{k}} \left(\frac{2\pi}{\beta \omega_{s}\left(\mathbf{k}, \beta \right)  }\right)  \right] ^{2}.
\end{equation}
It should be emphasized that the dispersion laws of elementary effective fields~$\omega_{s}\left( \mathbf{k}, \beta\right)$ depend on the parameters~ $\mu_{s},\ \varkappa_{s}$ of these fields, on the parameters $\gamma_{s}$ of the fields-atoms interactions and on temperature $T=1/\beta$ of the system.

Using the known relationship between the partition function~$Z$ of a system and its average energy~$E$
\begin{equation}\label{en}
E = -\frac{\partial }{\partial \beta}\left( \ln Z\right), 
\end{equation}
we find
\begin{equation}\label{ener}
E = \frac{3}{2}N T + 2 \sum_{s, \mathbf{k}} T -  2 \sum_{s, \mathbf{k}}  \frac{c^{2}}{2\omega_{s}^{2}\left( k,\beta\right) }\left( \frac{n \gamma_{s}^{2}}{2 \varkappa_{s}}\right)  .
\end{equation}
The second therm  the right-hand side of this formula contains infinitely many identical terms and is equal to infinity.  The sum of the series in the last term on the right side of this formula is also infinite. Therefore,
\begin{equation}\label{infty}
 E = \infty.
\end{equation}
The resulting infinity of auxiliary field energy is analogous to the paradox associated with the radiation of a heated body, known since the late 19th century as the ultraviolet catastrophe. In 1900, this paradox was resolved by Planck through the introduction of the quantum hypothesis.

\section{Quantum statistics of elementary effective fields}

Let us move from the classical description of the elementary effective fields ~$\psi_{s}\left(\mathbf{k} \right) $, $\chi_{s}\left(\mathbf{k} \right) $ to the quantum one. We assume that these fields are bosonic. Since the mass parameters~$\mu_{s}$ in the dispersion law~\eqref{dispers} are nonzero, we will call them \textit{massive oscillators}. Their energy spectrum is characterized by a continuous vector ~$\mathbf{k}$ and nonnegative integer quantum occupation numbers $n_{s}=0, 1, 2, 3, \ldots$:
\begin{equation}\label{quant}
	E_{s}\left(n_{s}, \mathbf{k} \right)  = \left( n_{s}+\frac{1}{2}\right) \hbar\omega_{s}\left( \mathbf{k}, \beta\right). 
\end{equation} 
This implies a relationship between infinitesimal changes in $k$ and $E_{n}\left(k \right) $:
\begin{equation}\label{dE-dk}
	dE_{n}^{s}\left( k, \beta\right) =\left(n_{s} + \frac{1}{2} \right) \frac{c\hbar k}{\sqrt{k^{2} + \mu_{s}^{2}  - \frac{n}{2 \varkappa_{s}} \beta\gamma_{s}^{2}}} dk.
\end{equation}

The number of states $N_{s}$ of the auxiliary field $\varphi_{s}\left(\mathbf{ k }\right) $ with  $\left|\mathbf{k} \right| \leq k$  is
\begin{equation}\label{Ns}
	N_{s} = \frac{V}{\left( 2\pi\right)^{3}} \, \frac{4}{3}\pi k^{3}
\end{equation} 
This implies that the number of states~$dN_{s}\left( k\right) $ in the layer between~$k$ and $k+dk$ 
\begin{equation}\label{dNs}
	dN_{s}\left( k\right)  = \frac{V}{\left(2\pi \right)^{3} }\, 4\pi k^{2}\,dk.
\end{equation}

The dispersion law of massive oscillators is as follows
\begin{equation}\label{Es}
	E_{s}\left( \mathbf{k}, \beta\right) =\hbar\, \omega_{s} \left( \mathbf{k}, \beta\right) = \hbar c \sqrt{k^{2} +\mu_{s}^{2}  - \frac{n}{2 \varkappa_{s}} \beta\gamma_{s}^{2} }.
\end{equation}

Let us express $dE_{s}(\mathbf{k})$ in terms of $dk$:
\begin{equation}\label{dEs-k}
	d\left(\hbar \omega_{s}\left(\mathbf{k}, \beta \right)  \right) = \hbar c \frac{k}{\sqrt{k^{2} + \mu_{s}^{2}  - \frac{n}{2 \varkappa_{s}} \beta\gamma_{s}^{2}   }}\, dk.
\end{equation}

The partition function of an individual harmonic~$\omega_{s}\left( \mathbf{ k} \right) $ has the known form
\begin{equation}\label{Z-s(omega)}
	Z_{s}\left(\mathbf{k}, \beta \right) = \sum\limits_{n=0}^{\infty} e^{-\beta \hbar \omega_{s}\left(\mathbf{k}, \beta \right)\left[ n + \frac{1}{2}\right]  } = \frac{e^{-\frac{\beta\hbar \omega_{s}\left( \mathbf{k}, \beta \right) }{2}}}{1-e^{-\beta\hbar\omega_{s}\left( \mathbf{k}, \beta\right) }}.
\end{equation}

From here we find the average energy of this harmonic
\begin{equation}\label{E-harm}
	E_{s}\left(\mathbf{k}, \beta \right) = \left[ \frac{\hbar \omega_{s}\left(\mathbf{k}, \beta\right)}{2} + \frac{\hbar \omega_{s}\left(\mathbf{k}, \beta \right)}{e^{\beta\hbar \omega_{s}\left(\mathbf{k}, \beta \right)} -1}\right] \left( 1 - \frac{c^{2}}{2\omega_{s}^{2}\left(\mathbf{k},\beta  \right) }\, \frac{n\beta\gamma_{s}^{2}}{2\varkappa_{s}} \right).
\end{equation}
Let's find the total energy of the $s$-th elementary effective field:
\begin{equation}\label{U-s}
\begin{array}{r}
{\displaystyle U_{s}\left( \beta\right)  = \frac{V}{\left(2\pi \right)^{3} } \int\limits_{0}^{\infty} \left[ \frac{\hbar \omega_{s}\left(\mathbf{k}, \beta\right)}{2} \right] \left( 1 - \frac{c^{2}}{2\omega_{s}^{2}\left(\mathbf{k},\beta  \right) }\, \frac{n\beta\gamma_{s}^{2}}{2\varkappa_{s}} \right) \, 4\pi k^{2}\,dk }\\
{\displaystyle + \frac{V}{\left(2\pi \right)^{3} } \int\limits_{0}^{\infty} \left[  \frac{\hbar \omega_{s}\left(\mathbf{k}, \beta \right)}{e^{\beta\hbar \omega_{s}\left(\mathbf{k}, \beta \right)} -1}\right] \left( 1 - \frac{c^{2}}{2\omega_{s}^{2}\left(\mathbf{k},\beta  \right) }\, \frac{n\beta\gamma_{s}^{2}}{2\varkappa_{s}} \right) \, 4\pi k^{2}\,dk. }
\end{array}	
\end{equation}
The first term on the right-hand side of this formula is a divergent integral and represents the infinite zero-point energy. The second term is the energy of the $s$-th elementary effective field, referenced from the energy of its zero-point oscillations:
\begin{equation}\label{W-beta}
W_{s}\left( \beta\right) =\frac{V}{\left(2\pi \right)^{3} } \int\limits_{0}^{\infty} \left[  \frac{\hbar \omega_{s}\left(\mathbf{k}, \beta \right)}{e^{\beta\hbar \omega_{s}\left(\mathbf{k}, \beta \right)} -1}\right] \left( 1 - \frac{1}{2 \left[  \frac{2T\varkappa_{s}\left(k^{2} + \mu_{s}^{2} \right)}{n\gamma_{s}^{2}}  - 1 \right]  }  \right) \, 4\pi k^{2}\,dk.
\end{equation}

Let us introduce a dimensionless integration variable  
\begin{equation}\label{Q}
	Q = \beta \hbar c k,
\end{equation}
as well as a parameter 
\begin{equation}\label{Ts}
T_{s} = \hbar c \mu_{s},
\end{equation}
which is a relativistic quantum characteristic of the elementary auxiliary field~$\varphi_{s}\left(\mathbf{r}, t \right) $, and find

\begin{equation}\label{sqrt-2}
\beta\hbar\omega_{s}\left(\mathbf{k}, \beta\right)  =\sqrt{Q^{2} + \left(\frac{T_{s}}{T} \right)^{2}\left[1 - \frac{n\gamma_{s}^{2}}{2\varkappa_{s}\hbar c  \mu_{s}^{3}} \left(\frac{T_{s}}{T} \right)  \right] }.
\end{equation}

A necessary condition for the applicability of this formula is
\begin{equation}\label{lim-T2}
\frac{T}{T_{s}} \geq \left\lbrace \frac{n\gamma_{s}^{2}}{2 \varkappa_{s}\hbar c\mu_{s}^{3} }\right\rbrace.
\end{equation}
Note that conditions~\eqref{lim-T2} and~\eqref{lim-T} are essentially identical, although they differ in form: variable~$\tau_{s}= T/T_{s}$ in~\eqref{lim-T2} contains the dimensionless temperature for the  $s$-th effective elementary field~\eqref{H(s)}. 

\begin{equation}\label{int-Q}
\begin{array}{r}
{\displaystyle  W_{s}\left(T \right)  =  \frac{V T^{4}}{\left( 2\pi\right)^{4} \left(\hbar c \right)^{3} } \int\limits_{0}^{\infty} \left[ \frac{\sqrt{Q^{2} + \left(\frac{T_{s}}{T} \right)^{2}\left[1 - \frac{n\gamma_{s}^{2}}{2\varkappa_{s}\hbar c  \mu_{s}^{3}} \left(\frac{T_{s}}{T} \right)  \right] }}{e^{\sqrt{Q^{2} + \left(\frac{T_{s}}{T} \right)^{2}\left[1 - \frac{n\gamma_{s}^{2}}{2\varkappa_{s}\hbar c  \mu_{s}^{3}} \left(\frac{T_{s}}{T} \right)  \right] } }-1}  \right] }\\
{\displaystyle \times \left(  { 1 - \frac{1}{2 \left\lbrace  \frac{2\varkappa_{s}\hbar c \mu_{s}^{3}}{n\gamma_{s}^{2}} \left(\frac{T}{T_{s}} \right)\left[\left(\frac{T}{T_{s}} \right)^{2} Q^{2} +1\right]   - 1 \right\rbrace  }    }  \right) Q^{2}\, dQ.    }	
\end{array}	
\end{equation}

This expression for the energy of the $s$-th elementary effective field consists of two factors, the first of which is universal and proportional $T^{4}$, and the second is individual and depends on the $\varkappa_{s}$, $\mu_{s}$, $\gamma_{s}$ parameters. 

The applicability of this formula is limited by the condition~\eqref{lim-T2}.
At high temperatures ~$T_{s}/T\to 0$ the second factor is equal to the constant~$\pi^{4}/15$, which, as expected, leads to the Stefan-Boltzmann law
\begin{equation}\label{Stefan}
W_{s}\left( T\right) \propto T^{4}.
\end{equation}

Let us consider the integral contained in formula~\eqref{int-Q}:
\begin{equation}\label{Int}
\begin{array}{r}
{\displaystyle f\left(\alpha_{s}, \tau_{s} \right) = \frac{15}{\pi^{4}} \int\limits_{0}^{\infty} \left[ \frac{\sqrt{Q^{2} + \frac{1}{\tau_{s}^{2}} \left[1 - \frac{\alpha_{s}}{\tau_{s}}  \right] } }{e^{\sqrt{Q^{2} + \frac{1}{\tau_{s}^{2}} \left[1 - \frac{\alpha_{s}}{\tau_{s}}  \right] } }-1}  \right] }\\
{\displaystyle \times \left(  { 1 - \frac{1}{2 \left\lbrace   \frac{\tau_{s}}{\alpha_{s} } \left[\tau_{s}^{2}  Q^{2} +1\right]   - 1 \right\rbrace  }    }  \right) Q^{2}\, dQ ,}
\end{array}
\end{equation}
where $\tau_{s}= \frac{T}{T_{s}}$ is dimensionless temperature, and 
\begin{equation}\label{alpha-s}
\alpha_{s} = \frac{n\gamma_{s}^{2}}{2\varkappa_{s}\hbar c \mu_{s}^{3}}
\end{equation}
is a dimensionless parameter consisting of the characteristics $\mu_{s}$, $\varkappa_{s}$ of the $s$-th elementary auxiliary field and the field-atoms coupling constant $\gamma_{s}$. 

According to formula~\eqref{Action}, the parameters $\varkappa_{s}$ are related to the energy of the free auxiliary fields $\varphi_{s}(\mathbf{r}, t)$, the parameters $\mu_{s}$ are determined by the singular points of the Fourier transform $\tilde{v}\left(k^{2} \right)$ of the corresponding static interatomic potential.
The factor~$({15}/{\pi^{4}})$ is introduced to normalize this integral to $1$ in the high temperature limit $\tau_{s}\gg 1$.

The graphs of the functions~$f\left(\alpha_{s}, \tau_{s}\right) $ for some particular values the $\alpha_{s}$ are shown in Fig.~1 and Fig.~2. 
\begin{figure}[h]
	\centering
	\includegraphics[width=1.2\linewidth]{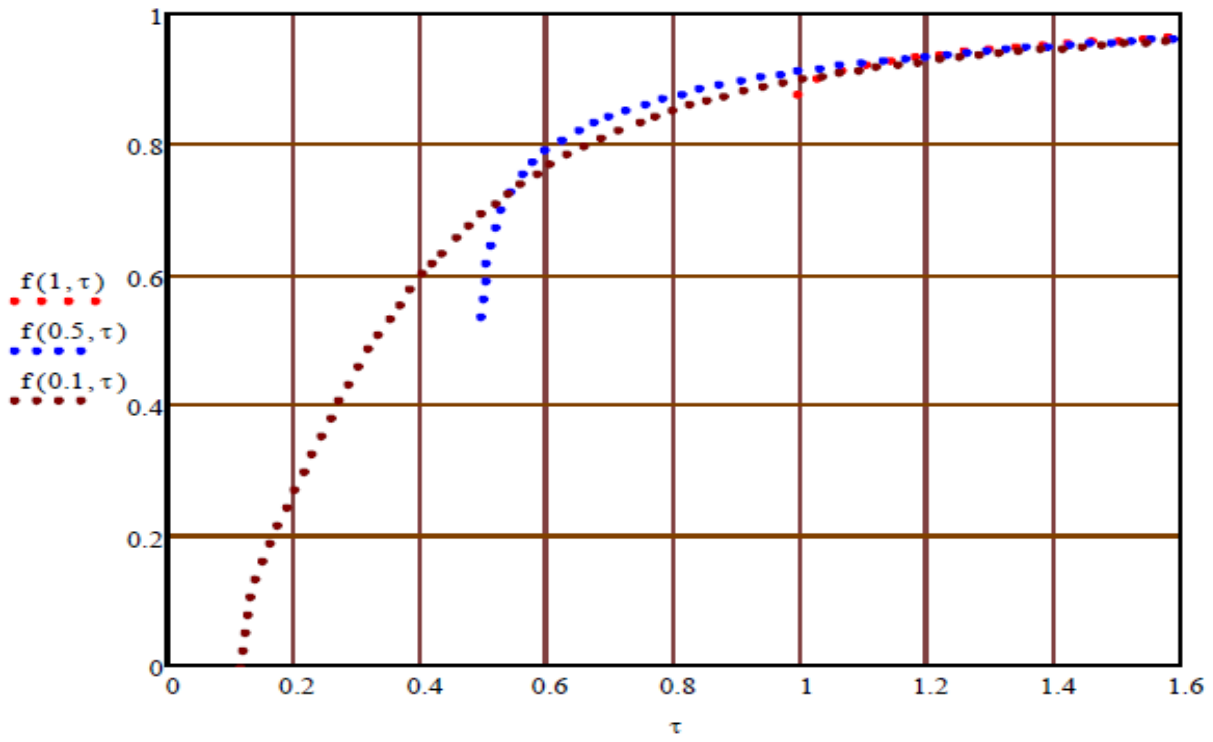}
	\caption{Graphs of functions $f(T/T_{s} )$ for $\alpha_{s} = 1.0, 0.5, 0.1$, respectively.}
	\label{fig:Fig-01}
\end{figure}

\begin{figure}[h]
	\centering
	\includegraphics[width=1.0\linewidth]{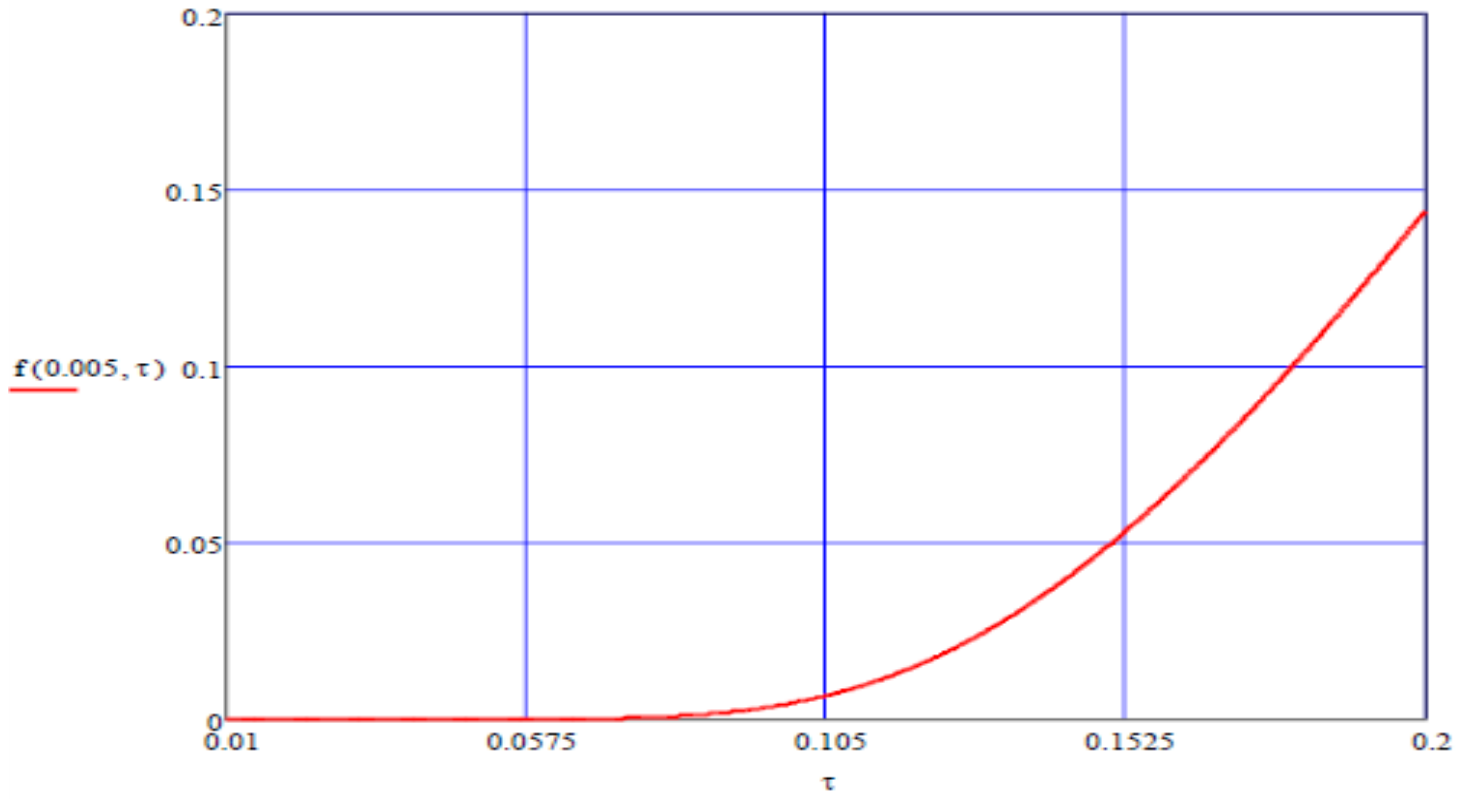}
	\caption{Graph of the function $f(\tau_{s} )$ in the low-temperature region ($\tau_{s}\ll 1$) in the case of a weak atom-field coupling constant ($\alpha_{s} = 0.005 \ll 1$).}
	\label{fig:fig-02}
\end{figure}
In accordance with the condition~\eqref{lim-T2}, the domain of definition of each of these functions is $\tau_{s} >\alpha_{s}$. Therefore, the boundary of the domain of definition of the function~$f\left(\alpha_{s}, \tau_{s}\right)$ point~$T_{\mathrm{crit}} = \alpha_{s}\, T_{s}$ is a singular point.

Thus, the explicit form of the function $f\left(\alpha_{s}, \tau_{s} \right) $ of the $s$-th effective elementary auxiliary field is completely determined by its dimensionless parameter $\alpha_{s}$.

\section{Critical point analysis}
The critical point corresponding to the $s$-th auxiliary field is the boundary point of the domain of definition of the function~$f\left(\alpha_{s}, \tau\right) $ as $\tau \to \alpha_{s}+0$.

Having expression~\eqref{int-Q} for the energy of the s-th effective auxiliary field, we can find the contribution of this field to the heat capacity of the system
\begin{equation}\label{capac}
C_{V} = \frac{4V T^{3}}{\left( 2\pi\right)^{4} \left(\hbar c \right)^{3} }f\left(\alpha_{s}, \tau_{s} \right) + \frac{V T^{4}}{\left( 2\pi\right)^{4} \left(\hbar c \right)^{3} }\, \frac{1}{T_{s}}\, \frac{\partial f\left(\alpha_{s}, \tau_{s}  \right)}{\partial \tau_{s}}.
\end{equation}

The functions $\frac{4V T^{3}}{\left( 2\pi\right)^{4} \left(\hbar c \right)^{3} }$, $f\left(\alpha_{s}, \tau\right)$, $\frac{V T^{4}}{\left( 2\pi\right)^{4} \left(\hbar c \right)^{3} } $ do not contain any singularities, unlike function
\begin{equation}
\frac{\partial f\left(\alpha_{s}, \tau  \right)}{\partial \tau} = 	f_{2}\left(\alpha_{s}, \tau\right),
\end{equation}
the qualitative form of which for particular values of the dimensionless $\alpha_{s}$ is presented in Fig.~3. 

\begin{figure}
	\centering
	\includegraphics[width=1.0\linewidth]{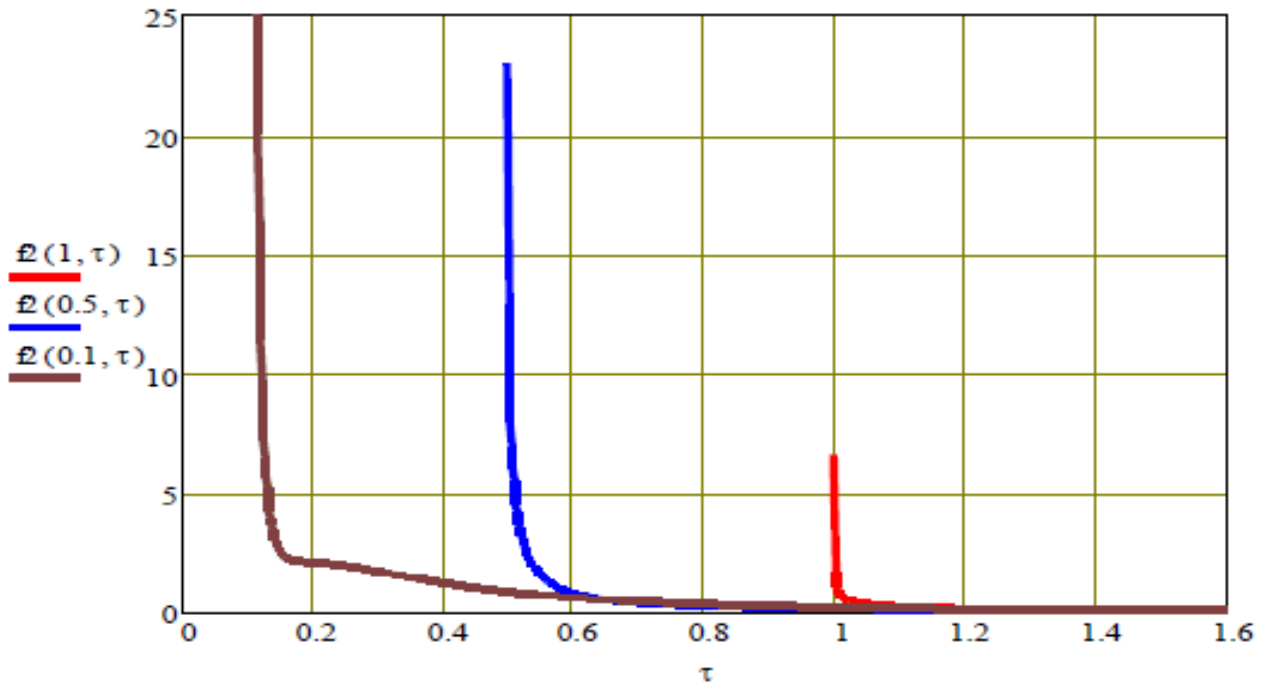}
	\caption{Qualitative form of functions $f_{2}\left(\alpha_{s}, \tau \right) $ for some particular values of $\alpha_{s}$}
	\label{fig:fig-03}
\end{figure}

Note that in all the presented cases there is a singularity in the heat capacity at $\tau = \alpha_{s}+0$. This indicates the existence of a phase transition within the framework of relativistic quantum statistical thermodynamics.

\section{Conclusion and prospects}
The main results of this work are as follows.
\begin{enumerate}
	\item The system of interacting atoms is represented as a union of two subsystems, one of which is a system of atoms, and the other is an auxiliary scalar covariant field, which is equivalent to a given static interatomic potential of a general form only in the non-relativistic approximation.
	
	\item It is shown that the auxiliary field is a superposition of Klein-Gordon fields, the mass parameters~$\mu_{s}$ of which are uniquely determined by the singular points of the Fourier transform of the static interatomic potential.
	
	\item An expression for the relativistic Hamiltonian of a system of interacting atoms is derived, incorporating both atomic and field degrees of freedom. The general form of the relativistic Hamiltonian of a system of interacting atoms is established. 
	
	\item An exact analytical calculation of the classical relativistic partition function of a system consisting of atoms and auxiliary fields has been performed.
	
	\item It has been established that, within the framework of relativistic statistical thermodynamics, atom-field interactions are equivalent to the renormalization of the parameters of auxiliary fields. As a result, the relativistic Hamiltonian of a system of interacting atoms can be represented as the sum of the Hamiltonian of free atoms and the Hamiltonian of a renormalized auxiliary field.
	
	\item It has been established that the divergence of the system's energy within the framework of classical relativistic statistical thermodynamics (analogous to the ultraviolet catastrophe in the theory of thermal radiation) is eliminated by quantizing the auxiliary fields.
	
	\item The existence of critical temperatures has been established.
	
	\item An interpretation of the process of establishing thermal equilibrium in a system of interacting atoms is proposed as a process of energy exchange between the auxiliary field and the atoms.

\end{enumerate}

\section*{Acknowledgements}

I am grateful to N.~P.~Alekseeva, Ya.~I.~Granovsky, M.~A.~Zakharov, and V.~V.~Zubkov for fruitful discussions.

\end{document}